\documentclass[a4paper]{article}
\usepackage{INTERSPEECH2022}
\usepackage[utf8]{inputenc} 
\usepackage[T1]{fontenc}    
\usepackage{hyperref}       
\usepackage{url}            
\usepackage{booktabs}       
\usepackage{amsfonts}       
\usepackage{nicefrac}       
\usepackage{microtype}      
\usepackage{xcolor}         
\usepackage{graphicx}
\usepackage{multicol}
\usepackage{multirow}
\usepackage{color, colortbl}
\usepackage{amsbsy}
\usepackage{array, makecell} %
\usepackage{boldline} 
\usepackage{hhline}
\usepackage[numbib]{tocbibind}
\usepackage[numbers,sort&compress]{natbib}
\usepackage[labelfont=bf]{caption}
\newcolumntype{P}[1]{>{\centering\arraybackslash}p{#1}}
\newcolumntype{M}[1]{>{\centering\arraybackslash}m{#1}}
\newcolumntype{N}{@{}m{0pt}@{}}

\definecolor{LightGray}{gray}{0.9}

\title{Directed Speech Separation for Automatic Speech
Recognition of Long-form Conversational Speech}
\name{Rohit Paturi, Sundararajan Srinivasan, Katrin Kirchhoff, Daniel Garcia Romero}
\address{Amazon AWS AI}
\email{paturi@amazon.com, sundarsr@amazon.com, katrinki@amazon.com, dgromero@amazon.com}

\begin{document}

\maketitle

\begin{abstract}
Many of the recent advances in speech separation are primarily aimed at synthetic mixtures of short audio utterances with high degrees of overlap. Most of these approaches need an additional stitching step to stitch the separated speech chunks for long form audio. Since most of the approaches involve Permutation Invariant training (PIT), the order of separated speech chunks is nondeterministic and leads to difficulty in accurately stitching homogenous speaker chunks for downstream tasks like Automatic Speech Recognition (ASR). Also, most of these models are trained with synthetic mixtures and do not generalize to real conversational data. In this paper, we propose a speaker conditioned separator trained on speaker embeddings extracted directly from the mixed signal using an over-clustering based approach. This model naturally regulates the order of the separated chunks without the need for an additional stitching step. We also introduce a data sampling strategy with real and synthetic mixtures which generalizes well to real conversation speech. With this model and data sampling technique, we show significant improvements in speaker-attributed word error rate (SA-WER) on Hub5 data.
\end{abstract}
\section{Introduction}

\begin{figure*}[htp]
  \centering
  \includegraphics[width=1\linewidth]{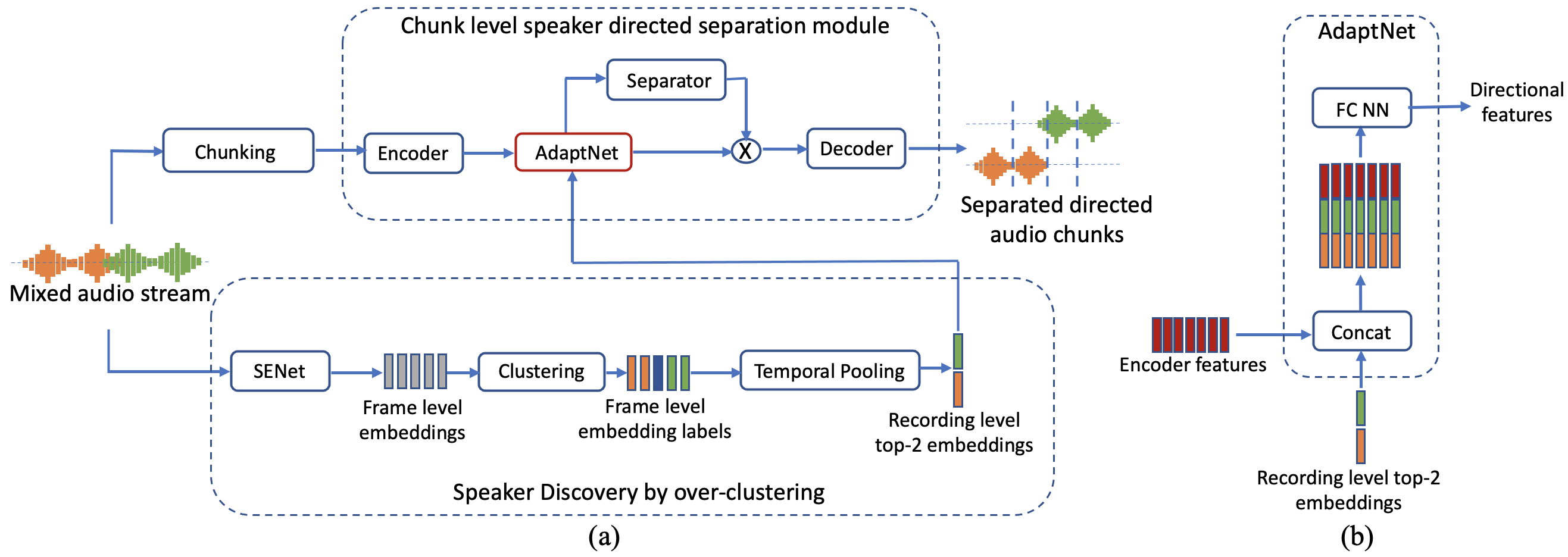}
  \vspace{-\baselineskip}
  \caption{(a) Overall architecture of the Directed Speech Separation for 2 speaker use-case, (b) Structure of the AdaptNet block.}
  \vspace{-1\baselineskip}
\end{figure*}

Despite the recent advances in Automatic speech recognition (ASR), multi-speaker scenarios still pose a significant challenge to ASR systems \cite{1yoshioka2018recognizing,2barker2018fifth,3kanda2018hitachi} because of the difficulty of attending to the target speech signal from other interfering speech signals. One approach to recognize multi-speaker overlapped speech is by end-to-end speaker-attributed automatic speech recognition (SA-ASR) systems \cite{4kanda2020joint,5kanda2020investigation,6kanda2021end,7raj2022continuous,8chang2021hypothesis} which jointly model speaker identification and speech recognition for monaural overlapped speech. Though these systems have shown promise in recognizing multi-talker speech, when more downstream tasks (like emotion recognition, speech diarization, etc.) from overlapped speech conversations are needed, every task needs to be re-trained in this framework, reducing modularity. Also, these are shown to not generalize well to long form audio \cite{8chang2021hypothesis}.
     
The other approach is to perform robust speech separation, which can then be a common frontend for all tasks and this is the approach we consider in this paper. Monaural speech separation has recently witnessed a rapid progress with the advent of supervised neural networks \cite{9wang2018supervised,10hersheydeep,11kolbaek2017multitalker,12luo2018speaker,13wang2018alternative} in the time-frequency domain and end-to-end time domain approaches \cite{14luo2019conv,15luo2020dual,16chen2020dual,17subakan2021attention}. One set of approaches leverage speaker information to improve the separation performance. These use either pre-enrolled speaker utterances to perform target speaker separation \cite{22zmolikova2021auxiliary,23wang2019voicefilter,24vzmolikova2019speakerbeam} or extract speaker from preliminary separation \cite{18zeghidour2021wavesplit,19nachmani2020voice,20wang2021dual,21byun2021monaural} to improve the speaker agnostic separation performance.
     
Most of these approaches above are trained with a PIT loss \cite{11kolbaek2017multitalker} leading to a nondeterministic ordering of separated channels. In order to apply these to long audio recordings, an explicit stitching step is needed to stitch the separated chunks to form the long homogenous speaker channels. A common stitching mechanism compares similarity between overlapping regions of adjacent chunks \cite{28han2020continuous,29chen2020continuous} to determine the correct chunk permutation to be stitched. But, this can be error-prone as one wrongly stitched chunk can lead to error propagation throughout the subsequent chunks and is sensitive to the separation quality of every chunk. Also, studies \cite{25cosentino2020librimix,26kadiouglu2020empirical,27menne2019analysis} report that, even though these separation models have consistently advanced the state of the art on some of the popular synthetic datasets in the field like wsj0-mix \cite{10hersheydeep} and LibriMix \cite{25cosentino2020librimix}, the ability to generalize to speech coming from real conversation settings in terms of ASR performance has not been achieved. 

In this paper, we propose a speaker conditioned 2-speaker speech separation model for conversational telephone speech (CTS) without the need for pre-enrolled utterances and doesn’t require PIT loss as the separated channels are directed by the order of the speakers fed into to the model. An over-clustering based approach is used to find speaker embeddings robust to speech overlaps which serve as inputs for the Directed Speech Separation (DSS) module. To train our system, we propose a data sampling strategy leveraging both synthetic read speech and in-domain real conversational datasets. We demonstrate that applying this DSS as a frontend to ASR on long-form audio is superior to stitching separation outputs with a PIT loss trained model, with an SI-SDR improvement of 10dB on CALLHOME American English \cite{39LDC} and SA-WER \cite{4kanda2020joint} improvement of 30\% on Hub5 dataset \cite{42LDC}.

\section{Related Work}

Two related approaches that don’t rely on pre-enrolled speaker utterances are Wavesplit \cite{18zeghidour2021wavesplit} and Continuous speech separation using speaker inventory (CSSUSI) for long recording \cite{28han2020continuous}. 

Wavesplit performs preliminary separation and speaker embedding extraction followed by clustering to extract speaker centroids. These are then used to condition the separation stack. Wavesplit has been explored for only short fully overlapping utterances and is complex to train due to multiple stages of separation and speaker stack involved. Our approach differs from Wavesplit as it doesn’t need any preliminary separation and can make use of a strong pre-trained speaker embedding network reducing the complexity of the system substantially.

CSSUSI directly extracts embeddings from mixed speech and forms a speaker inventory to condition the separation network. The separation network operates in time-frequency domain and is trained with PIT loss and hence, the order of the separated chunks is nondeterministic. A stitching mechanism using overlap similarity with adjacent chunks is used to stitch back the separated chunks. Our approach differs from CSSUSI mainly by conditioning a more robust end-to-end time domain separator network without the need for an additional stitching mechanism. 

\section{Directed Speech Separation}
\label{SS}

In this section, we introduce the components of the directed speech separation (DSS) system. It mainly comprises of two modules: a speaker discovery module robust to speaker overlaps and a speaker conditioned directed separation module. 

\subsection{Robust Speaker Discovery by Over-Clustering}

The speaker discovery module is used to discover and extract embeddings of the constituent speakers of an audio mixture by taking advantage of the large number of non-overlapping speech regions in multi-talker conversational scenarios. A pretrained speaker embedding network (SENet) extracts frame level speaker embeddings   \(\{\boldsymbol{f_i}\}_{i=1}^F, \boldsymbol{f_i} \in \mathbb{R}^{1\times K}\)  , where F is the number of frames in the recording. These embeddings are clustered using spectral clustering with maximum eigen gap \cite{32wang2018speaker} for detecting the number of clusters C with additional constraints such that \(N\leq C\leq M\) where N is the number of speakers to be separated and M is the maximum detected clusters, where \(M > N\). Thus, we over-cluster the embeddings by setting these constraints and show in §4.4 that separation performance is insensitive to the value of M as long as \(M > N\) and that over clustering the embeddings produces cleaner speaker clusters by attributing overlapped or noisy/background speech to additional clusters. Once the frame level embeddings \(\{\boldsymbol{f_i}\}_{i=1}^F\) are partitioned into clusters \(I_1,..., I_C\) of sorted cardinalities \(n_1,...,n_C\), such that \(n_1>...>n_C\) and \(n_1+...+n_C=F\), the top-N utterance level embeddings \(\{\boldsymbol{z_j}\}_{j=1}^N, \boldsymbol{z_j} \in \mathbb{R}^{1\times K}\) are computed by temporally pooling the frame level embeddings of the corresponding N clusters. In this paper, we use a simple mean pooling to produce recording level embeddings 
\vspace{-0.5\baselineskip}
\[ \boldsymbol{z_j}=\frac {1}{n_j}\sum_{\boldsymbol{f_i}\in I_j}\boldsymbol{f_i} \vspace{-0.5\baselineskip}\]

\noindent These embeddings are used by the separation network to condition and direct the network as outlined in §3.2. This speaker discovery module is pretrained and is frozen during training and inference. 

\subsection{Speaker Directed Separation Network}
The speaker directed separation module separates audio at the chunk level without the need for restitching the separated chunks in order to separate a long utterance. The encoder, separator and decoder architectures are based on ConvTasNet \cite{14luo2019conv} architecture in this paper but can be based on any of the more recent transformer architectures \cite{16chen2020dual,17subakan2021attention}. The inputs to this module are the audio chunk x, the recording level embeddings \(\{\boldsymbol{z_j}\}_{j=1}^N\) and the outputs are the separated chunk level waveforms \(\{\boldsymbol{\hat{s_j}}\}_{j=1}^N\) .The audio chunk x is divided into overlapping segments of length L, represented by \(\{\boldsymbol{x_k}\}_{i=1}^T, \boldsymbol{x_k} \in \mathbb{R}^{1\times L}\), where T denotes the total number of encoder frames in the input chunk. \(\boldsymbol{x_k}\) is transformed into a E dimensional encoder representation, \(\{\boldsymbol{e_k}\}_{k=1}^T, e_k \in \mathbb{R}^{1 \times E}\) by a 1-D convolution operation:
\vspace{-0.5\baselineskip}
\[\boldsymbol{e_k}= \mathcal{H}(\boldsymbol{x_k} \boldsymbol{U})\vspace{-0.5\baselineskip}\]
where \(\boldsymbol{U}\in \mathbb{R}^{L\times E}\) contains E encoder basis functions with length L each and \(\mathcal{H}(\cdot)\) is the ReLU non-linear function. We introduce an adaptation network (AdaptNet) shown in Fig. 1(b), which concatenates the N recording level speaker embeddings \(\{\boldsymbol{z_j}\}_{j=1}^N\) to each frame of the encoder features \(\boldsymbol{e_k}\) to form the intermediate directional features \(\{\boldsymbol{a_k}\}_{k=1}^T, \boldsymbol{a_k} \in \mathbb{R}^{1 \times A}\), such that
\vspace{-0.5\baselineskip}
       \[\boldsymbol{a_k}=concat(\boldsymbol{e_k},concat(\boldsymbol{z_1},...,\boldsymbol{z_N})),\] 
where \(A=E+N\times K\). The intermediate directional features are transformed by a fully connected neural network to form the D-dimensional directional features \(\{\boldsymbol{d_k}\}_{k=1}^T, \boldsymbol{d_k} \in \mathbb{R}^{1 \times D}\)
\vspace{-0.5\baselineskip}
\[\boldsymbol{d_k}= \mathcal{H}(\boldsymbol{a_k} \boldsymbol{W})\vspace{-0.5\baselineskip}\]
where \(\boldsymbol{W}\in \mathbb{R}^{A \times D}\) is the AdaptNet weight matrix, and \(\mathcal{H}(\cdot)\) is the ReLU non-linear function. The separator consists of stacked dilated temporal convolutional networks \cite{14luo2019conv} and predicts a representation for each of the N sources by learning N masks \(\{\boldsymbol{m_j}\}_{j=1}^N, \boldsymbol{m_j} \in \mathbb{R}^{1 \times D}\) such that \(\boldsymbol{m_j}\in [0,1]\). The representation of each separated source \(\{\boldsymbol{t_{k,j}}\}_{j=1}^N, \boldsymbol{t_{k,j}} \in \mathbb{R}^{1 \times D}\) is calculated by applying the corresponding mask \(\boldsymbol{m_j}\) to the directional features \(\boldsymbol{d}\):
\vspace{-0.5\baselineskip}
\[\boldsymbol{t_{k,j}}=\boldsymbol{d_k} \odot \boldsymbol{m_j}\vspace{-0.5\baselineskip}\]
where \(\odot\) denotes element-wise multiplication. The waveform of each separated overlapping segment \(\boldsymbol{\hat{s}_{k,j}} \in \mathbb{R}^{1 \times L}\) is reconstructed by the decoder:
\vspace{-0.5\baselineskip}
\[\boldsymbol{\hat{s}_{k,j}}= \boldsymbol{t_{k,j}} \boldsymbol{V}\vspace{-0.5\baselineskip}\]
where \(\boldsymbol{V} \in \mathbb{R}^{L\times E}\) contains E decoder basis functions. The overlapping reconstructed segments are summed together to generate the separated chunk level sources \(\boldsymbol{\hat{s}_j}\). Since we condition the separator using speaker embeddings, we train the network to minimize the negative scale invariant signal to distortion ratio (SI-SDR) \cite{36le2019sdr} between separated sources \(\{\boldsymbol{\hat{s_j}}\}_{j=1}^N\) ordered consistently with the input speaker embeddings \(\{\boldsymbol{z_j}\}_{j=1}^N\) and the ground truth sources \(\{\boldsymbol{s_j}\}_{j=1}^N\) and thus, avoid the permutation problem. As we need to align \(\{\boldsymbol{\hat{z_j}}\}_{j=1}^N\) to the corresponding \(\{\boldsymbol{s_j}\}_{j=1}^N\) for training the separator network, we leverage Hungarian algorithm of md-eval tool\cite{34SCTK} to provide the best alignment.

\section{Experiments}
\label{exps}
\subsection{Datasets}
This work aims at evaluating speech separation as a front-end for ASR for long form CTS data. In order to train and evaluate the separation model, we use the two channel Fisher \cite{40LDC,41LDC} and CALLHOME American English (CHAE) \cite{39LDC} datasets respectively. Fisher contains ~2000 hours and CHAE contains \(\sim60\) hours of largely two speaker conversations, available in separate channels. Two speaker conversations were filtered from Fisher using the call and speaker metadata. We generate mixture data by mixing both the channels into a single channel and labeling the individual channels as the ground truths for speech separation. Along with these CTS datasets, we also use synthetic fully overlapping Libri2Mix \cite{25cosentino2020librimix} and wsj0-mix \cite{10hersheydeep} to train the models using the data selection strategy outlined in §4.3. We use the default CHAE dev and test sets \cite{39LDC} for finding the best sampling coefficient (§4.3) and separation quality evaluation respectively. For ASR evaluations, we use HUB5 2000 English data \cite{42LDC} which is a \(\sim11\) hours CTS dataset used for evaluating CTS ASR systems. Popular benchmarks \cite{30xiong2016achieving,31saon2017english} on HUB5 consider the telephone channels separately. In this work, we mix them to create single-channel HUB5 and report ASR performance before and after speech separation. The average duration of a conversation in CHAE is \(\sim30\) minutes and is $\sim$10 minutes for Fisher and HUB5 datasets. All CTS datasets in this work have a sampling rate of 8KHz and the synthetic mixes were downsampled to 8KHz.

\begin{figure}[t]
  \centering
  \includegraphics[width=1\columnwidth]{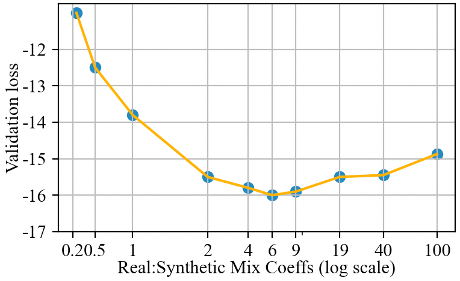}
  \vspace{-1.5\baselineskip}
  \caption{Sampling coefficient parameter search on dev set}
  \label{fig:Sampling Coefficient}
  \vspace{-1\baselineskip}
\end{figure}

\subsection{Implementation Details}
We solve for two-speaker separation use-case in this paper but this can be extended to more speakers by training the separator for multi speaker use-case similar to the base separation architectures in \cite{14luo2019conv,15luo2020dual,16chen2020dual,17subakan2021attention}. The SENet for this work is modelled using the ResNet34 architecture and is pretrained with a combination of classification and metric loss \cite{33chung2020defence} with 12k speakers and 4k hours of CTS data. The frame duration for embedding extraction is 0.5 seconds and the embedding dimension is 512. The extracted embeddings were augmented with Gaussian noise \cite{18zeghidour2021wavesplit} in addition to the implicit noise due to overlapping speech and clustering errors for training the separator. At train time, we also randomly flip the pooled embeddings along with the target separation signals. This is dynamically applied to 50\% of the training samples and helps improve the generalization of the system. The separator in this work follows the best ConvTasNet architecture in \cite{14luo2019conv} and is trained with 8s chunks. We train the separator with Adam optimizer with a batch size of 32 and learning rate of 1e-3 for 100 epochs. We set M to a large value of 6 (analysis on test set in §4.4) to account for any noisier recordings in training and N is 2 for the two-speaker separation use-case. We call the ConvTasNet trained with PIT loss as undirected speech separator (USS) as it produces outputs in a nondeterministic order. For the conversational English ASR system, we use the pretrained Aspire model from Kaldi \cite{35povey2011kaldi}.

\subsection{Data Sampling Strategy}
Previous works \cite{25cosentino2020librimix,26kadiouglu2020empirical,27menne2019analysis} have reported subpar separation performance on realistic datasets when trained with the fully overlapping synthetic datasets. In addition, we also observe that relying only on real conversational data is not optimal as the amount of single speaker regions outweighs the amount of overlapping speaker regions by a large margin (approximately 10:1), causing skewed data for training the separation models. So, we propose a data sampling strategy (RealSynMix) which leverages both synthetic mixes (LibriMix) along with CTS data (Fisher). During training, we sample the fully overlapping synthetic data and real conversational data parameterized by a sampling coefficient which defines the ratio of real to synthetic utterances to be sampled in each batch and is treated as a hyperparameter, learnt by optimizing the separation performance on the CHAE dev-set. We choose a sampling coefficient of 6 (6 parts of the Fisher sampled with 1 part of LibriMix) for our experiments, as it has the lowest negative SI-SDR from Figure ~\ref{fig:Sampling Coefficient}.

\begin{table}[t]
  \caption{SI-SDR (dB) of a ConvTasNet (USS) model trained and evaluated on different simulated and real datasets. CHAE is evaluated at the chunk level}
  \vspace{-0.2\baselineskip}
  \label{tab:data sampling table}
  \centering
    \begin{tabular}{P{14mm}P{9mm}P{9mm}P{17mm}P{9mm}N} 
    \hline\hline
        Train Data   & wsj0mix  & LibriMix & SparseLibriMix & CHAE \\ \hline\hline
 wsj0mix  & 15.8   & 8.3 & 8.1 & -2.2 \\
        LibriMix       & 14.2   & 14.5 & 22 & 4.2 \\
        Fisher & 9.4   & 12 & 20 & 14.1 \\
       RealSynMix & 14.2 & 14.4 & 21.8 & \textbf{15.5} \\ \hline
    \end{tabular}
    \vspace{-1.6\baselineskip}
\end{table}

We also compare the performance of this sampling strategy on commonly used separation datasets wsj0mix, LibriMix and SparseLibriMix \cite{25cosentino2020librimix} in Table ~\ref{tab:data sampling table}. For these experiments, SparseLibriMix has been generated with 9\% overlap to simulate the overlap in CHAE and the SI-SDRs are evaluated at the chunk level, where the chunk size was 8s. From Table 1, we can see that the performance of the model trained with RealSynMix significantly outperforms the performance of wsj0mix and LibriMix trained models on the real CHAE dataset while also performing well on the synthetic mixes. It also outperforms the Fisher only trained model on the simulated datasets as well as real CHAE. Though SparseLibriMix was also generated with the same overlap ratio as CHAE, the SDRs on SparseLibriMix with LibriMix trained models being much better than CHAE shows that the simulated sparse datasets derived from audiobooks don’t fully capture the conversational structure and dynamics of CTS data well enough. Also, the synthetic mixes are derived from read speech whereas conversation speech is the typical use-case for speech separation and the ASR system that follows.
\begin{table}[t]
\centering
\caption{Chunk and Recording level SI-SDR (dB) on CHAE dataset at different recording durations to highlight the efficiency of the DSS system over USS system on long-form audio.} \label{tab:SI-SDR}
\vspace{-0.5\baselineskip}
\begin{tabular}{P{6mm}P{11mm}P{10mm}P{5mm}P{5mm}P{5mm}P{5mm}}
\hline\hline
\multirow{2}{*}{Model} & \multirow{2}{*}{\makecell{Max\\Clusters M}} & \multirow{2}{*}{\makecell{Chunk\\Level}} & \multicolumn{4}{c}{Recording level at durations} \\
\cline{4-7}
& & & 20s & 100s & 300s & 600s \\
\hline\hline
USS & - & 15.5 & 15.1 & 12.8 & 10.5 & 6.7 \\
\hline
\multirow{4}{*}{DSS}& 2 & 14.5 & 14.4 & 14.3 & 14.3 & 14.4 \\
& 3	& 16.5 & 16.5 &	16.4 &	16.5 & 16.5 \\
& \textbf{4} & \textbf{16.6} & \textbf{16.6} & \textbf{16.4} & \textbf{16.5} & \textbf{16.6} \\
& 5	& 16.6 & 16.6 & 16.4 & 16.5	& 16.6 \\
\hline
\end{tabular}
\vspace{-1\baselineskip}
\end{table}
\subsection{Directed speech separation}

To compare the performance of the DSS system with the USS system, we evaluate the chunk level SI-SDR on the held-out test subset of CHAE. To evaluate the directedness of the system, we evaluate the recording level SI-SDR for different durations of audio on the CHAE test set. For the recording level evaluations, the outputs of USS are stitched with adjacent overlapping chunk similarity as in \cite{29chen2020continuous}. The chunk size during inference is 8s with no overlap for DSS and has an overlap of 4s for USS with stitching.

From Table ~\ref{tab:SI-SDR}, we see that not only does the DSS system improve the chunk level separation quality, it also remains consistent across different durations of the recordings. On the other hand, the USS system performance degrades as the duration of recording increases. This is mainly due to error propagation following an erroneous stitched chunk as the stitching relies only on the adjacent chunks. These erroneous stitches can happen frequently based on the separation quality and as the number of chunks increase with the recording duration. 

We also analyze the effect of number of clusters on the separation quality in Table ~\ref{tab:SI-SDR} and show over clustering ($M>N$) improves the separation quality due to cleaner speaker clusters. It can be observed that the separation quality significantly improves for $M=3$ compared to $M=2$. This is due to some of the noisy and overlapping speech being attributed to the 3rd cluster for $M=3$, producing cleaner and more robust top-2 speaker embeddings. The separation quality is almost identical once the number of clusters is not fewer than the number of speakers, i.e. $M>2$. The separation quality slightly improves for $M=4$ compared to $M=3$ as few noisier utterances are assigned an extra cluster for the noisy/overlap regions. The maximum number of detected clusters using max eigen gap across all CHAE utterances was 4 and hence the results for $M>4$ are exactly identical.

Finally, we evaluate the ASR performance of both the systems on the HUB5 dataset in Table ~\ref{tab:WER table}. We pass single channel HUB5 through the separators followed by the ASR system to get the WERs of the separated audio. We also pass the single channel HUB5 directly through ASR without any speech separation frontend to get WERs for unseparated audio. The oracle SA-WER is obtained by passing the oracle speaker channels of the original multi-channel HUB5 independently through the ASR system. We also report the SA-WER on the non-overlap (non-ovl) regions, i.e. single speaker regions to compare the separation performance in areas of no speech overlap. ASCLite \cite{38fiscus2006multiple} which can align multiple hypotheses against multiple reference transcriptions, is used to calculate the SA-WERs.

\begin{table}[t]
\centering
\caption{SA-WER (\%) on HUB5 (CH and Switchboard subsets). Full, Non-ovl are SA-WERs of full utterance and non-ovl regions}\label{tab:WER table}
\vspace{-0.5\baselineskip}
\begin{tabular}{P{20mm}P{6mm}P{12mm}P{6mm}P{12mm}}
\hline\hline
\multirow{3}{*}{Model} & \multicolumn{4}{c}{Hub5 Subsets}\\
\cline{2-5}
& \multicolumn{2}{c}{Callhome} & \multicolumn{2}{c}{Switchboard} \\
\cline{2-5}
& Full & Non-ovl & Full & Non-ovl\\
\hline\hline
\rowcolor{LightGray}
\begin{tabular}{@{}c@{}}None \\ (unseparated)\end{tabular} & 26.3 & 20.7 & 25.5 & 13.3\\
\hline
{Oracle channels}	& 18.4 & 17.9 & 10.6 & 9.7 \\
USS & 52 & 48.3 & 46.2 & 42.8\\
DSS & \textbf{23.0} & \textbf{19.2} & \textbf{14.6} & \textbf{12.8}\\
\hline
\end{tabular}
\vspace{-1\baselineskip}
\end{table}
The DSS model improves the SA-WER on HUB5 by 24\% relative (13\% and 43\% on the CallHome (CH) and Switchboard (SWBD) subsets respectively) compared to unseparated HUB5 data which shows the clear advantage of the DSS frontend in conversational ASR. We see that the USS system fails heavily on both subsets in terms of SA-WER as well, similar to the SI-SDR numbers on long recordings. Another important observation is that the SA-WER of non-overlapping regions with the DSS frontend is also better than unseparated non-overlap (non-ovl) SA-WER though these regions comprise of just single speaker speech. This can be attributed to the ASR (mainly language models) having better context by separating the adjacent overlapping speech regions. Finally, there is still a good difference between the oracle SA-WER and the DSS SA-WER, suggesting that there is still room for improvement for the long form directed speech separation model.

\section{Conclusion}
In this work, we introduced a speaker conditioned directed speech separation (DSS) model for long form real conversational telephone speech (CTS). This uses an over-clustering based approach to extract robust speaker embeddings without the need for pre-enrolled utterances. This not only naturally directs and stitches the separated short chunks in the order of the extracted speaker embeddings, but also improves the separation quality of the short chunks. In addition, we highlighted drawbacks of using some of the popular simulated datasets for training a CTS separation model. We solved this by proposing a data sampling strategy that combines the benefits of both real and synthetic datasets which shows significant improvements on the speech separation quality for CTS data when compared to the synthetic datasets or real datasets alone. With the DSS model, we achieved with an SI-SDR improvement of 1dB on short form and 10dB on long form CALLHOME American English and a SA-WER improvement of \(\sim30\%\) on Hub5 dataset compared to the PIT based undirected speech separation (USS) model.

Future work will focus on scaling the system to a variable number of speakers, designing a block-online system instead of an offline system and improving the separation performance with stronger conditioning techniques and base separator architectures using Transformer networks \cite{16chen2020dual,17subakan2021attention}. 
\bibliographystyle{ieeetr}
{\eightpt
\bibliography{citations}}

\end{document}